 \documentstyle[11pt,aaspp4]{article}

\begin{document}

\title{Interferometric observation of the L483 molecular core}

\author{Y.-S. Park\altaffilmark{1,2,3}, J.-F. Panis\altaffilmark{4}, 
N. Ohashi\altaffilmark{1}, M. Choi\altaffilmark{1}, and 
Y.C. Minh\altaffilmark{2}   }

\altaffiltext{1}{Institute of Astronomy and Astrophysics, Academia Sinica, 
P.O. Box 1-87, Nankang, Taipei, Taiwan 115, R.O.C.}
\altaffiltext{2}{Korea Astronomy Observatory, San 36-1, Whaam-dong, Yusung, 
Taejon, 305-348, Korea}
\altaffiltext{3}{Astronomy Program, SEES, Seoul National University, San 
56-1, Shillim-dong, Kwanak-gu, Seoul, 151-742, Korea}
\altaffiltext{4}{Radioastronomie, \'Ecole Normal Sup\'erieure, 24 rue Lhomond,
75231, Paris cedex 05, France}

\begin{abstract}

We present the aperture synthesis imaging of C$_3$H$_2$ 2$_{12}$--$1_{01}$ 
and HCO$^+$ 1--0 lines and of continuum emission at $\lambda=3.4$ mm 
toward the class 0 young stellar object, IRAS 18148-0440 in the L483 
molecular cloud.
The continuum emission is detected at the IRAS position at a level of 16 mJy, 
indicating a compact source with a mass of $\sim 0.13 M_{\odot}$.
The C$_3$H$_2$ delineates an envelope near the IRAS position, with a 
size of $\sim$ 3000 AU $\times$ 2000 AU in R.A. and Dec. directions, 
respectively.  A velocity gradient detected only along the axis of 
outflow can be explained in terms of free-fall motion of the envelope. 
The HCO$^+$ line wing extends up to the velocity of $\pm 6$ km s$^{-1}$ 
relative to the systemic one, and the high velocity material shows a
symmetric bipolarity and extends over $2'$ or 0.1 pc along the east-west 
direction.  It is found that the outflow material is clumpy and the opening 
angle is widest for the slowest moving component.
The core component of the HCO$^+$ 1--0 line exhibits an anti-infall 
asymmetry not only in interferometric but also in single dish observations.  
It is attributable to the slow isotropic outward motion of gas between 
the flattened envelope and the collimated outflow. 

\end{abstract}


\section{Introduction}

Young stellar objects (YSOs) in the early phase of star formation 
continuously aggregates material from parent dense molecular cores.
This process seems to be followed by the formation of disk and outflows,  
and protostars come into existence by dispersing debris around them 
(\cite{kit97}).
This scenario is generally accepted at least for low mass stars.
The classification or evolutionary sequence of YSOs from class 0 
to III, mainly based on their spectral energy distribution, has been 
established and became a framework to which subsequent studies should 
refer (\cite{and93}).

However, one needs to find direct observational evidence of inward 
motion leading to a drastic density enhancement of central source which 
is prerequisite for the star formation.
The method of diagnosing the collapse motion had been developed earlier 
(\cite{leu77}; \cite{ang87}), but it was first detected only in the 1990's 
(\cite{zho93}).
The method is based on the idea that, in the presence of the infall motion 
of which magnitude is not so large as that of intrinsic (turbulent + thermal) 
motion, an opaque line has a self-absorption with the blue peak stronger 
than the red one, while an optically thin line with a single peak is located
between the two peaks of the opaque line.
Since class 0 and I objects are believed to be in main accretion phase, 
prompt attention is paid to them in an attempt to understand the earliest 
phase of star forming process.  
Many class 0 and I sources have been surveyed initially with single dish 
telescopes mainly for the detection of infall signatures, resulting  in a 
statistical preference for inward motion rather than outward one for 
these objects (\cite{gre97}; \cite{mar97}). 
These surveys also served as a basis for detailed studies of higher angular 
resolution.  Imaging with high spatial resolution is particularly 
important since the star formation process takes place at a smaller scale.
Only interferometric observations enable us to separate envelopes, disks, 
and outflow lobes and to disentangle motions associated with them.
Choi et al. (1999) carried out a survey of 9 class 0 sources using 
the Berkeley-Illinois-Maryland-Association (BIMA) interferometer, and 
Hogerheijde et al. (1999) on the YSOs in the Serpens molecular cloud.  
Detailed investigation has been undertaken for a limited number of YSOs 
as well (\cite{oha97}; \cite{yan97}).
Eventually all YSOs will have to be observed with interferometers in 
a variety of transitions.

As a step toward these high resolution studies, we conducted an 
interferometric observation of L483.  The L483 is one of the so-called 
Myers' cores containing an embedded infrared source, IRAS 18148-0440, at 
a distance of 200 pc.  Its size is $\sim 1'$ from the observations of 
HC$_3$N (\cite{ful93}) and C$^{18}$O (\cite{ful97}).
A CO outflow with a good collimation was found in the east-west direction, 
inclined by $\sim 45^{\circ}$ to the sky plane, and an infrared lobe also 
lies in the axis of molecular outflow (\cite{par91}; \cite{ful95}).
Recently, Hatchell, Fuller, \& Ladd (1999) and Buckle, Hatchell, 
\& Fuller (1999) examined the temperature variation in the outflow lobe 
and the shock structure of jet embedded in the lobe, respectively. 
However, the kinematics of the central condensation remains poorly 
understood, since the line asymmetry of optically thick transitions, known 
as a probe of internal motions of cores, differs from molecule to molecule. 
Mardones et al. (1997) and Mardones (1998) obtained CS 2--1, H$_2$CO 
$2_{12}$--$1_{11}$, and N$_2$H$^+$ 1--0 lines with $20''\sim 30''$ beams 
and mapped their spatial distributions.  They found that the CS and 
H$_2$CO lines are self-reversed and the blue peak is slightly brighter 
than the red.  
On the other hand, Gregersen et al. (1997) show that the red peak is 
twice as strong as the blue one in the HCO$^+$ 3--2 transition in 
a $\sim 20''$ beam.
The three hyperfine components of HCN 1--0 again exhibit the usual blue
asymmetry with a $\sim 60''$ beam: a single peak in the blue side with a 
shoulder in the red of the thinnest component ($F$=0--1) changes 
into a deep self-absorption with a stronger blue peak of the 
thickest one ($F$=2--1) (\cite{par99}).
A recent VLA observation of NH$_3$ indicates an inward motion down to 
$\sim 10''$ scale (\cite{ful00}).
All these features suggest diverse velocity fields in this source, 
and this could be understood much better by aperture synthesis imaging.

In section 2, we briefly describe the observation and the data reduction.  
Results of our interferometric observation are presented in section 3.  
In section 4, we discuss kinematics of the envelope and outflow as well 
as molecular abundance.  In the final section, we summarize our results.

\section{Observation and data reduction}

Since the HCO$^+$ molecule is most peculiar in line shape when compared 
with other molecules in single dish observations, the HCO$^+$ molecule 
is the one to be explored with a prime importance in this observational 
study.  The HCO$^+$ 1--0 transition has a critical density of 
$\sim 2.5 \times 10^5$ cm$^{-3}$.  The HCO$^+$ molecule may not be 
depleted in an envelope since there is no chemical reaction route to 
consume it (\cite{raw96}; \cite{van98}).  Moreover, it often delineates 
well outflow lobes (\cite{hog98}; \cite{hog99}).  Thus it may be a good  
tracer of the density and motion of the envelope and the outflow.  
However, since the transition may be opaque, we selected  
C$_3$H$_2$ 2$_{12}$--$1_{01}$ as a complementary probe to cold gas in 
the envelope.
The C$_3$H$_2$ molecule has been extensively surveyed toward numerous
Galactic objects (\cite{mad98}; \cite{ben98}).
The single-peaked line profile of C$_3$H$_2$ toward L483 obtained from a
single dish observation suggests that the transition is not so optically 
thick (\cite{mye95}).
Furthermore, in L1157 it is found to trace an envelope mainly and is 
insensitive to an outflow or shocked region (\cite{bac97}).  
Thus, if a disk-like envelope is embedded in the core of L483, we may be 
able to discern, by using C$_3$H$_2$, rotation and/or radial 
(inward/outward) motions which are not contaminated by the outflow.


The observation of the L483 was carried out with the 10 elements BIMA 
interferometer in its C configuration in October and November 1998 in 
the transitions of HCO$^+$ 1--0 (89.188523 GHz) and C$_3$H$_2$ 
2$_{12}$--1$_{01}$ (85.338905 GHz).  
We configured correlator for HCN 1--0 line (88.631847 GHz) as well, 
but the HCN data were discarded because of interference in the IF band.  
System temperatures were typically $150\sim 500$ K during the observation. 
Projected baselines range from 1.7 k$\lambda$ to 23 k$\lambda$, and
phase center was the position of IRAS 18148-0440, 
$\alpha{\rm(2000)}= 18^h 17^m 29.\!\!^s83$ and 
$\delta{\rm(2000)}=-4^{\circ} 39' 38.\!\!''33$.
The phase and pass band were calibrated using 1743-038 and 3C273, 
respectively.  We also made observations of Uranus for flux calibration 
and the flux of 1743-038 was $1.8 \sim 2.0$ Jy at the time of observation.
The IF window with the band width of 12.5 MHz was fed into the 256 
channels of the correlator for each transition, resulting in the velocity 
resolutions of 0.164 km s$^{-1}$ and 0.172 km s$^{-1}$ in the HCO$^+$ 
and C$_3$H$_2$ transitions, respectively.  Continua in both upper and 
lower side bands were simultaneously observed with a 800 MHz bandwidth in 
total.

We reduced the data using the MIRIAD package. The data were flagged and 
Fourier transformed with natural weighting.  The resulting dirty map 
was CLEANed and restored with a Gaussian beam of a FWHM size of
$14'' \times 10''$ at position angle $5.\!\!^{\circ}5$ measured from 
North to East.

In addition, in January 1999, one point observation of HCO$^+$ 1--0 was 
undertaken toward the IRAS source using the radome-enclosed 14 meter 
telescope (beam size = $60''$) of Taeduk Radio Astronomy Observatory, 
Korea.  
The SIS receiver was used for the frontend and the system temperature was
400 K at the time of observation.  The backend was an autocorrelation 
spectrometer with a 20 KHz or 0.067 km s$^{-1}$ resolution.

\section{Results}

\subsection{Continuum emission at $\lambda=3.4$ mm}

Fig.~1 represents the continuum emission detected around the position of 
IRAS 18148-0440.  There is diffuse emission in the NE--SW direction.  
The continuum source is barely resolved in the direction of major beam 
axis, but extended in the E-W direction. 
By deconvolving the beam, we roughly estimate the size of condensation, 
2000 AU and 3000 AU in the N-S and E-W directions, respectively.  
The estimated total flux is 16 mJy, which is consistent with spectral 
energy distribution measured from centimeter to near infrared 
(\cite{ful95}).  Using a mass absorption coefficient, 
$\kappa = 0.01 ({1.3 {\rm mm} \over \lambda})^{1.5}$
cm$^2$ g$^{-1}$ (\cite{mot98}), 
we obtain the mass of 0.13 M$_{\odot}$ for the condensation.  

\subsection{C$_3$H$_2$ line emission}

We illustrate the distribution of C$_3$H$_2$ in Fig.~2.  The line is 
single-peaked with the FWHM width of 0.32 km s$^{-1}$ and centered at 
the systemic velocity of 5.4 km s$^{-1}$, which is in accordance with the
single dish observations of optically thin N$_2$H$^+$ and C$_3$H$_2$ lines
(\cite{mye95}).
The emission peak at $\approx$ ($-9'', -3''$) is displaced by $6''$ to 
the west of the continuum emission.  The C$_3$H$_2$ emission is extended 
from North to South and its half-power contour larger than the synthesized 
beam is almost circular.  From the figure, we note that the C$_3$H$_2$ 
emission delineates the envelope around the protostar fairly well.  
The single dish observation at high spatial resolution in the C$^{18}$O 
3--2 transition revealed the structure that is similar, but slightly 
elongated in the N-S direction (\cite{ful97}).  

In order to examine the kinematics of the envelope, we plot channel
maps in Fig.~3, where there are mainly three components.  The main 
component at $\approx$($-10'', 0''$) with $v\approx 5.4$ km s$^{-1}$  
contains most of the mass of the envelope.  The remaining ones are
at $\approx$($5'', 0''$) with $v\approx 5.0$ km s$^{-1}$ and at 
$\approx$($15'', 20''$) with $v\approx 6.0$ km s$^{-1}$.
Extended features in the N--S direction in the figure frame of 
5.41 km s$^{-1}$ may result from side lobe effect.  
The 5.0 and 6.0 km s$^{-1}$ components are minor fragments of 
the envelope.  The systematic displacement of peak positions of the 
main component along with the LSR velocity suggests a velocity gradient 
which is seen more clearly in the position-velocity diagrams of Fig.~4.  
Along the outflow axis of ${\rm P.A.}=95^{\circ}$ (see section 3.3) 
through the C$_3$H$_2$ emission peak at ($-9'', -3''$), is a uniform 
velocity gradient of the main component, amounting to 
$\sim 0.5$ km s$^{-1}$ over $40''$ or 13 km s$^{-1}$ pc$^{-1}$.  
It is found that line wing is of little importance as expected in 
section 2 and the line {\em itself} is gradually shifted.  The figure 
also shows that the 5.0 km s$^{-1}$ feature is isolated indeed. 
On the other hand, the position-velocity diagram across the outflow 
axis shows little sign of systematic motion.  We are not confident 
with the current resolution that the component at 
$\Delta \delta \approx -20 ''$ with 5.6 km s$^{-1}$ is a discrete one.  

A plausible explanation of those velocity shifts would be that the 
envelope has a flattened structure and the velocity gradient represents 
a gas infall motion toward the central source, whereas the existence of 
rotation is uncertain (\cite{oha97}).  
However, one has to be careful since the sense of the velocity shift 
of infalling flat disk or envelope is always the same as that of outflow 
motion.  
For instance, the shift of C$_3$H$_2$ line shown in L1228 results
from the fragments of core entrained to an outflow (\cite{taf97}).
It is difficult to figure out which one is dominant for L483, given the 
C$_3$H$_2$ data only.  However, the velocity shift by outflow is less 
likely for L483, since the HCO$^+$ line suggests that there is expanding 
motion as a part of outflow around the envelope and the velocity field is 
almost isotropic (see section 4.2).  If the C$_3$H$_2$ line is affected by 
this motion, there will be little velocity gradient.  Fuller and Wooten 
(2000) recently found the inward motion by carrying out the VLA observation 
using the NH$_3$ inversion lines.

The infall motion of L483 inferred from the position-velocity map of 
Fig.~4 is different from a free-fall motion of $v(r)=\sqrt{2 G M_* / r}$ 
toward a central star with a mass of $M_*$ which has usually been invoked 
for, e.g., L1527, HH111, and other YSOs (\cite{oha97}; \cite{yan97}; 
\cite{mom98}).
The velocity gradient is rather linear, suggestive of a collapse of a 
(pressure-free) gas sphere with a uniform density.
We would argue that it is not an accretion toward the central source, 
but a global contraction of the envelope.  This is consistent with 
the current understanding of the class 0 objects (\cite{bac96}). 
Our interpretation, however, may be subject to the spatial and spectral 
resolutions of observation.  Further observations with higher resolutions 
are necessary to elucidate the kinematics and structure of the envelope.

\subsection{HCO$^+$ emission}

In Fig.~5, we plot the distribution of the red and blue HCO$^+$ 1--0 wing 
emission.  The red wing is integrated from 6.5 to 11.0 km s$^{-1}$, while
the blue wing from $-1.5$ to 4.5 km s$^{-1}$.
It is found that the bipolar outflow is well collimated and the peaks 
of the two outflow lobes are separated each other by $\sim 50''$.  
The shape of the outflow is similar to those found from CO 2--1 
(\cite{par91}), from CO 3--2 (\cite{ful95}), and from CO 4--3 
(\cite{hat99}).
The outflow lies close to the R.A. axis (P.A.$\approx 95^{\circ}$).  
The different amounts of extinction towards the lobes in the IR band 
suggest an inclination of the outflow of $\sim 45^{\circ}$ to the sky 
plane (\cite{ful95}). 

In order to see details of the outflow, we display channel maps of HCO$^+$ 
in Fig.~6.  It is found that there are at least three discrete components 
in the red shifted outflows: the 5.9 km s$^{-1}$ component around 
$\approx$($35'', 10''$), the 6.1 km s$^{-1}$ component centered at 
$\approx$($10'', -5''$), and the 7.2 km s$^{-1}$ component near 
$\approx$($35'', 0''$).  
The 7.2 km s$^{-1}$ component spans a wide range of velocity from 6.7
to 10 km s$^{-1}$.  The 5.9 km s$^{-1}$ component is much more widely 
distributed than the others which are relatively compact.
The strong emission around ($-11'', 0''$) in the velocity range of 
$5.9 \sim 6.4$ km s$^{-1}$ represents the red side of line core 
component coming from the envelope.  There is little emission in the 
vicinity of line center ($v=5.2 \sim 5.7$ km s$^{-1}$) over the 
whole field of view, due to heavy self-absorption.  
The emission on the blue side ($v=4.5 \sim 5.1$ km s$^{-1}$) of the 
systemic component is absent, contrary to the one in the red side 
(which will be discussed in the next section).
Interestingly, the blue outflow lobe also seems to have three discrete 
components and to share similar kinematics as the red lobe.  One can
find the 3.3 km s$^{-1}$ component farthest from the systemic velocity, 
located around $\approx$($-10'', 5''$) whose velocity range is wide from 
1.4 to 4.3 km s$^{-1}$.
The other two components have velocities of $\approx$ 4.4 and 5.0 
km s$^{-1}$ which reside at $\approx$($5'', 5''$) and ($15'', -10''$), 
respectively.  The 5.0 km s$^{-1}$ component closest to the rest velocity 
covers the largest area as the 5.9 km s$^{-1}$ one does in the red lobe. 

The fact that the individual velocity components can be identified 
and grouped into three pairs supports the idea of symmetric and episodic 
mass ejections.  
It may be claimed from the number of pairs that the mass loss phenomena 
have taken place at least three times in the past.
There must be an ambient cloud or a cavity wall with velocity close to the 
systemic one which embeds these different velocity components, but this 
is probably resolved out, since its distribution would be extended.  
The shortest baseline mentioned in section 2 suggests that we are blind to 
structures larger than $\sim 100''$.
The clumpiness and intermittency of the outflow is rather common in the 
case of extreme high velocity outflows (\cite{bac96}).  It seems that 
standard or low velocity outflows exhibit such features as well 
(cf. \cite{hog98}; \cite{gom99}).

Furthermore, we note that the component whose velocity is closer to the 
systemic one occupies larger area.  This can be evidence that an opening 
angle of the outflow becomes wider as the YSO evolves (\cite{bac96}), 
provided that the ejected material is accelerated after the ejection from 
around zero velocity.  
If this assumption holds, then the high velocity component will be 
located far away from the driving source.  
Actually the velocity roughly proportional to the distance away from the 
source is a reasonable approximation to the velocity field of the 
well-collimated flow, as can be seen in HH211, NGC1333-IRAS2, NGC2071, 
and NGC2264G (\cite{mas93}; \cite{gue99}; \cite{san94}; \cite{che92}; 
\cite{fic98}).
It is approximately true for the L483 as well, which is shown in the 
position-velocity diagram of Fig.~7. 
The velocity increase together with the distance from the source could 
also be made by other mechanisms like the simultaneous outburst of material 
with a wide range of velocities.  However, this may not be applicable to 
the well-collimated outflows with narrow throats. 

On the other hand, one can imagine that the opening angle has been kept 
wide from the beginning, but the fastest mass ejection has taken place 
preferentially normal to the flat disk or envelope.  Although this is not
the case for the general class 0 objects, this possibility can not be 
ruled out.
The slowly moving material with wider opening angle could also be regarded 
as a remnant of a pre-existing shell which was slowly expanding and is 
swept up by the fast jet.  All these invoke detailed understanding of the 
outflow and hence should be tested by further observation.

\section{Discussion}

\subsection{Abundance estimate of HCO$^+$ in the outflow}

The line strength of HCO$^+$ in the wings indicates that the molecule may 
be enhanced in the outflow region.  We estimate the column density of 
HCO$^+$ and CO in the outflow lobe in order to derive the relative abundance 
of HCO$^+$ to CO.
Since the line wings are as wide as 5 km s$^{-1}$ along any line of sights,
the condition of the {\em large velocity gradient} is applicable.  Then we 
can use an expression by Goldreich \& Kwan (1974):
\begin{equation}
\tau={8 \pi^3 \mu^2 \over 3 h }{R \over V} 
n_0 (1-e^{-T_0 / T_{\rm ex}}),
\end{equation}
where $\mu(=3.4 {\rm debye})$ is the dipole moment of HCO$^+$, $R/V$ is 
the velocity gradient, $n_0$ is the level population in the level $J=0$, 
$T_{\rm ex}$ is the excitation temperature, and $T_0(=h \nu_{10} / k$) 
is 4.2 K.  If we use an approximation, 
$(n_0+n_1+n_2+\cdots)/n_0 \equiv n_t/n_0 \simeq 2 k T_{\rm ex} / h \nu_{10}$,
\begin{equation}
\tau={2 \pi^3 \mu^2 \nu_{10} \over 3 k T_{\rm ex} V } N({\rm HCO}^+) 
(1-e^{-T_0 / T_{\rm ex}}),
\end{equation}
where the column density $N$(HCO$^+$) is equal to $2 R n_t$.

The $\tau$ is derived from 
\begin{equation}
T_{\rm b}=(J(T_{\rm ex})-J(2.7 K)) (1-e^{-\tau}),
\end{equation}
where $J(T)=T_0/(e^{T_0/T}-1)$.
Since $T_{\rm b}\approx 1$ K in the wings (see Fig.~8), $\tau \approx 0.15$ 
for an assumed $T_{\rm ex}$ of 10 K.  $N$(HCO$^+$) is then  
$1.4\times 10^{13}$ cm$^{-2}$ for $V=5$ km s$^{-1}$.
In the case that $T_{\rm ex}= 20$ K, $\tau \approx 0.051$ and 
$N({\rm HCO}^+) = 1.7\times 10^{13}$ cm$^{-2}$.

We derive the column density of CO in the outflow lobe in a similar way 
by using the $J$=2--1 transitions of both $^{12}$CO and $^{13}$CO in 
Hatchell, Fuller, \& Ladd (1999).  
The resulting CO column density is $N({\rm CO})=4.2\times 10^{17}$ cm$^{-2}$, 
where we adopted $\tau = 4.0$ and $T_{\rm ex}=25$ K.  Finally it is 
found that the abundance of HCO$^+$ relative to CO is 
$(3.3 \sim 4.0) \times 10^{-5}$.  
This is comparable to the value of $\sim 10^{-4}$ inferred either for the 
Orion extended ridge or for the TMC-1 region (\cite{irv87}).
Thus we find no significant abundance enhancement of HCO$^+$ in the 
outflow.  The line wings look rather strong simply due to the depression 
of the line core. 

\subsection{Anti-infall signature of the HCO$^+$ lines}


Line profiles of HCO$^+$ 1--0 toward the envelope are displayed in Fig.~8, 
where a spectrum obtained toward the IRAS source with the single dish 
telescope is shown together.
Emission from line core is mildly self-absorbed in the single dish 
observation, but suffers from severe self-absorption in the array 
observation, falling off close to zero.  The self-absorption of the 
line core is usually amplified in the interferometric observation, since 
the rest velocity component is most extended and the lack of visibility 
data at short spacing filters it out. 
Here it should be pointed out that the blue peak is weaker than the red 
one in the single dish observation and even almost missing in the 
interferometric one.
This asymmetry is consistent with those of higher transitions of HCO$^+$ 
(\cite{gre97}). 
All these observations strongly suggest that the stronger red peak is 
not an artifact caused by the limited UV coverage of interferometric 
observation, but an intrinsic property of the envelope of L483.  
The stronger red peak of optically thick line is generally known as 
an {\em anti-infall} signature, indicative of an outward motion. 
However, it contradicts the observations of HCN, H$_2$CO, and NH$_3$, all
implying the inward motion (\cite{par99}; \cite{mye95}; \cite{ful00}).

What gives rise to the anti-infall asymmetry of HCO$^+$?  In answering 
this question, however, we should not violate observational facts of 
HCN, H$_2$CO, and NH$_3$ supporting the collapse motion.
First, we can consider the influence of the outflow.
In addition to the self-absorption by the surface layer of the envelope,
the blue lobe of the outflow in front of the envelope may further 
obscure the blue part of line core emission.
However, since the outflow will have a higher excitation temperature than
the envelope (\cite{hat99}), it makes the emission from the envelope 
brighter irrespective of its optical depth: 
if the outflow is optically thin, the emission from the outflow should be 
added to that of envelope, while if the outflow is opaque, then the emission 
from outflow will replace it. 
Therefore the outflow can not explain the anti-infall asymmetry of the line 
core.  

The asymmetry seems to arise from the envelope itself or very close to it.  
The envelope or material in its vicinity may be really in a state of 
almost isotropic expansion. 
Here we recall that the material (traced by HCO$^+$) whose velocity is 
close to the systemic one moves outward with a wide opening angle, 
whereas the envelope (traced by C$_3$H$_2$) is collapsing.  
Thus one viable option may be a nearly isotropic outward motion of gas 
between the disk-like envelope and the collimated outflow, which is 
actually a part of the outflow.  
If this motion takes place over a large amount of volume around the 
flattened envelope, then we will have line profiles with the anti-infall 
signature.  
The detail of the line shape depends on the velocity field in this 
expanding volume, but in any case the line will have the red peak 
stronger than the blue (\cite{leu78}).  
To summarize, there is a highly collimated jet-like flow far from the 
central source, while there is a wide angle wind-like mass ejection at the 
base of the outflow.  Therefore the geometry and kinematics of the outflow 
of L483 seems to be in favor of the X-wind model as a low velocity version
(cf. \cite{bac96}; \cite{li96}).
The Cep A-HW2 is one of the recent examples showing similar line shapes 
as L483 (\cite{gom99}).  

Then one may have to look for reason why only the HCO$^+$ reflects the 
expansion.
This could be explained in terms of its abundance and excitation condition.  
The main route to form HCO$^+$ is the reaction between H$_3^+$ and CO.  The 
HCO$^+$ is unlikely to be depleted, since both species are abundant in the 
quiescent and relatively diffuse clouds (\cite{van98}; \cite{lan96}).  
As a consequence, although HCO$^+$ has a large critical density, it is 
collisionally excited easily at lower densities due to the line trapping.  
In this case, the critical density, $n_c$, should be modified as 
$A_{ji}/(C_{ji} \tau_{ji})$, where $\tau_{ji}$ is an optical depth, as 
already pointed out by Genzel (1991).  
It is also shown with a simple LVG analysis that HCO$^+$ 1--0 transition
requires the lowest effective density, $n_{eff}$, among CS, H$_2$CO, HCN, 
and HCO$^+$ to produce the line intensity of 1 K for a given column density 
per unit velocity interval of $\log (N / \Delta V) = 13.5$, where the 
column density, $N$, is given in cm$^{-2}$ and the line width, $\Delta V$, 
in km s$^{-1}$ (\cite{eva99}).
(At T$_k=10$ K, $n_{eff}=2.4 \times 10^3$ cm$^{-3}$ for HCO$^+$ 1--0, 
while $n_{eff}=1.8 \times 10^4$, $6.0 \times 10^4$, and 
$2.9 \times 10^4$ cm$^{-3}$, for CS 2--1, H$_2$CO 2$_{12}$--1$_{11}$, and 
HCN 1--0, respectively.)
Thus the expanding part of the envelope could be more effectively traced 
by the HCO$^+$ 1--0 transition than by other transitions.
Park, Kim, and Minh (1999) compared the line profiles of HCO$^+$ 3--2 \& 
4--3, H$_2$CO 2$_{12}$--1$_{11}$, CS 2--1, and HCN 1--0 of 9 class 0 and I 
objects, and found that the HCO$^+$ molecule seems to prefer the anti-infall
asymmetry than the HCN and possibly the CS does.  The preference is very 
marginal due to a small number of samples, but this could be indirect
evidence supporting the idea mentioned above.

\subsection{Position of the young stellar object in L483}

As shown in Fig.~1, the position of continuum peak coincides with that 
of IRAS 18148-0440 within $2'' \sim 3''$, while the peak of C$_3$H$_2$ 
emission is displaced to the west by $6''$ from it.
Although HCO$^+$ is severely saturated, the emission peak of the red part 
of the systemic component at $v\approx 5.9$ km s$^{-1}$ is roughly 
coincident with that of C$_3$H$_2$, as represented in Fig.~6.

Are the two sources really different condensations?  The error ellipse 
of the IRAS source position is as large as $35'' \times 8''$ with 
a position angle of $86^{\circ}$ (IRAS Point Source Catalogue).
However, it is found from the VLA observation that the 3.6 cm and 6 cm 
continuum source is detected exactly at the IRAS position and there is 
no emission at the position of the C$_3$H$_2$ emission peak. 
Although the VLA beam is $19''\times 9''$ and $10''\times 5''$ at 3.6 
cm and 6 cm, respectively, the accuracy in determining the peak position 
should be better than $1'' \sim 2''$ (\cite{ang97}; \cite{ang00}; 
\cite{bel00}).  The far-infrared ($100 \sim 190\mu$m) and submillimeter 
($450 \sim 1100\mu$m) continuum emission are also detected at the IRAS 
source position, but the beams are as large as
$\approx 50''$ and $\approx 20''$, respectively (\cite{ful95}; 
\cite{lad91}).
Recently Fuller and Wooten (2000) mapped the region at $\lambda = 450$ 
and $850 \mu m$ and found an elongated condensation at the IRAS position
with ${\rm P.A.}\approx 70^{\circ}$ whose (beam-convolved) FWHM size is 
$\approx 30'' \times 15''$.
Thus there may be an object at $\approx$ ($-9'', -3''$) which is unveiled 
only in millimeter line emission and possibly in the continuum from 
sub-millimeter to far-infrared.
However, it is more likely that the line emission of C$_3$H$_2$ and HCO$^+$ 
and the continuum emission emanate from the same condensation but that 
the molecular line emission is depressed around the dust condensation 
by some reasons.  One of the reasons may be that the gas is frozen out 
onto the grain in the cold and dense environment.  One can find such examples
in VLA 1623 and NGC 2024 (\cite{and93}).



\section{Summary}

The BIMA array observation has been undertaken toward the IRAS 18148-0440, 
a YSO embedded in the L483 molecular core.
An envelope with a beam-deconvolved size of 
$\sim 3000$ AU (R.A.) $\times$ 2000 AU (Dec.) and with a mass of
$\sim 0.13 M_{\odot}$ is found to collapse towards its center. 
The collapse motion is different from the Shu type ($v\propto r^{-1/2}$), 
but more likely linear with a velocity gradient of $\sim 13$ 
km s$^{-1}$ pc$^{-1}$.
The YSO seems to undergo main accretion phase without any conclusive 
signature of rotation.
The molecular outflow traced by HCO$^+$ 1--0 suggests that there have 
been successive (at least three times) mass loss events to date and that 
the recent ejection might take place with a wide opening angle compared 
with previous ones. 
The rest velocity component of the HCO$^+$ line is found to exhibit the
anti-infall signature, while the other transitions of density tracing 
molecules like CS, H$_2$CO, and HCN are not.
If the outflow is poorly collimated at its base, the outward motion 
directed radially would be pervasive in both sides of the flattened 
envelope and responsible for the HCO$^+$ 1--0 line biased to the red.
Due to the ease of collisional excitation and little chance of depletion, 
the HCO$^+$ 1--0 is the most likely transition with which one can sense 
the outward motion. 

\acknowledgments
We thank the anonymous referee for his careful reading of the manuscript.
YSP was partially supported by the BK21 program of Ministry of Education, 
Korea.
 

\clearpage

\figcaption{Map of continuum emission at $\lambda=3.4$ mm.  Contour 
levels are from $\pm 2 \sigma$ in steps of $1.5 \sigma$, where 
$1 \sigma = 0.8$ mJy beam$^{-1}$.  The position of IRAS 18148-0440 
or the phase center is marked with a cross at 
$\alpha{\rm(2000)}= 18^h 17^m 29.\!\!^s83$ 
and $\delta{\rm(2000)}=-4^{\circ} 39' 38.\!\!''33$.
The synthesized beam is shown in the bottom left.
\label{fig1}}

\figcaption{Map of the C$_3$H$_2$ intensity integrated from 4.5 to 6.5 
km s$^{-1}$.  Contour levels start at $\pm 2 \sigma$ with an 
increment of $1.5 \sigma$, where $1 \sigma = 0.1$ Jy beam$^{-1}$ km s$^{-1}$ 
\label{fig2}}

\figcaption{Channel maps of C$_3$H$_2$.  Contour levels are from 
$\pm 4 \sigma$ with an increment of $4 \sigma$, where 
$1 \sigma = 0.1$ Jy beam$^{-1}$.  
The LSR velocities are designated in the upper left corner of each 
figure frame.  
\label{fig3}}

\figcaption{Position-velocity diagram of C$_3$H$_2$ along the outflow 
axis (left) and perpendicular to it (right).  Contour levels are from 
$4 \sigma$ by an increment of $4 \sigma$, where $1 \sigma = 0.11$ K.  
Beam size and spectral resolution are indicated in the bottom left of 
each panel.
\label{fig4}}

\figcaption{Integrated intensity map of blue ({\rm solid contour}) 
and red ({\rm dashed contour}) wings of the HCO$^+$ line superimposed 
on the continuum map ({\rm grey}).  
The blue wing is integrated from $-1.5$ to 4.5 km s$^{-1}$ and the red 
wing from 6.5 to 11.0 km s$^{-1}$.  Both the lowest contour and the 
increment step are $ 5\sigma$ for both wings, where 
$1 \sigma = 0.1$ Jy beam$^{-1}$ km s$^{-1}$.
\label{fig5}} 

\figcaption{Channel maps of HCO$^+$.  Contour levels are from 
$\pm 5\sigma$ Jy beam$^{-1}$ by an increment of $5\sigma$ Jy beam$^{-1}$, 
where $1 \sigma = 0.1$ Jy beam$^{-1}$.  The LSR velocities are  
designated in the upper left corner of each figure frame. 
Note that 3 frames between the LSR velocities of 5.10 and 5.76 
km s$^{-1}$ are skipped since they show little emission due 
to the heavy self-absorption.
\label{fig6}}

\figcaption{Position-velocity diagram of HCO$^+$ along the outflow 
axis (${\rm P.A.}=95^{\circ}$) through the (0,0) position.  Contour 
levels are at $3\sigma$ intervals from $5\sigma $, where $1\sigma = 0.11$ K.
Beam size and spectral resolution are indicated in the bottom left. 
\label{fig7}}

\figcaption{Line profiles of HCO$^+$ from the array (left) and the 
single dish (right) observations towards the envelope.
Offsets from the IRAS source position are shown in the upper left  
corner of each panel for the interferometer observation.
 \label{fig8}}


\begin{thebibliography}{}

\bibitem[Andr\'e, Ward-Thompson, \& Barsony 1993]{and93} Andr\'e, P., 
Ward-Thompson, D., Barsony, M., 1993, ApJ, 406, 122


\bibitem[Anglada  et al. 1987]{ang87} Anglada, G., Rodr\'{\i}gez, L.F., 
Cant\'o, J., Estalella, R., L\'opez, R., 1987, A\&A, 186, 280

\bibitem[Anglada, Sep\'ulveda, \& G\'omez 1997]{ang97} Anglada, G., 
Sep\'ulveda, I., G\'omez, J.F., 1997, A\&AS, 121, 255

\bibitem[Anglada et al. 2000]{ang00} Anglada et al., 2000, in 
preparation

\bibitem[Bachiller 1996]{bac96} Bachiller, R., 1996, ARAA, 34, 111
 
\bibitem[Bachiller \& P\'erez Guit\'errez 1997]{bac97} Bachiller, R., 
P\'erez Guit\'errez, M., 1997, ApJ, 487, L93
 
\bibitem[Beltr\'an et al. 2000]{bel00} Beltr\'an et al., 2000, in preparation

\bibitem[Benson, Caselli, \& Myers 1998]{ben98} Benson, P.J., Caselli, P., 
Myers, P.C., 1998, ApJ, 506, 743

\bibitem[Buckle, Hatchell, \& Fuller 1999]{buc99} Buckle, J.V., Hatchell, J.,
Fuller, G.A., 1999, A\&A, 348, 584

\bibitem[Chernin \& Masson]{che92} Chernin, L.M., Masson, C.R., 1992, 
ApJL, 396, 35

\bibitem[Choi, Panis, \& Evans 1999]{cho99} Choi, M., Panis, J.-F., 
Evans, N.J.II, 1999, ApJ, 122, 519 


 

\bibitem[Evans 1999]{eva99} Evans, N.J.II, 1999, ARAA, 37, 311

\bibitem[Fich \& Lada 1998]{fic98} Fich, M., Lada, C.J., 1998, ApJS,
117, 147
 
\bibitem[Fuller et al. 1995]{ful95} Fuller, G.A., Lada, E.A., 
Masson, C.R., Myers, P.C., 1995, ApJ, 453, 754.
 
\bibitem[Fuller \& Ladd 1997]{ful97} Fuller, G.A., Ladd, E.F., 1997, 
IAU Symp. No. 182, Herbig-Haro Flows and the Birth of Low Mass Stars, 
eds. B. Reipurth and C. Bertout (Kluwer: Netherlands), p.495
 
\bibitem[Fuller \& Myers 1993]{ful93} Fuller, G.A., Myers, P.C., 1993, 
ApJ, 418, 273
 
\bibitem[Fuller \& Wooten 2000]{ful00} Fuller, G.A., Wooten, A., 2000, 
ApJ, in press 
 
\bibitem[Genzel 1991]{gen91} Genzel, R., 1991, Chemistry in Space, 
eds. J.M. Greenberg and V. Pirronello (Kluwer: Netherland), p. 123

\bibitem[Goldreich \& Kwan 1974]{gol74} Goldreich, P., Kwan, J., 1974,
ApJ, 189, 441

\bibitem[G\'omez et al. 1999]{gom99} G\'omez, J.F., Sargent, A.I., 
Torelles, J.M., Ho, P.T.P., Rodr\'{\i}guez, L.F., Cant\'o, J., Garay, G., 
1999, ApJ, 514, 287


\bibitem[Gregersen et al. 1997]{gre97} Gregersen, E.M., Evans, N.J.II, 
Zhou, S., Choi, M., 1997, ApJ, 484, 256
\bibitem[Gueth \& Guilloteau]{gue99} Gueth, F., Guilloteau, S., 1999,
A\&A, 343, 571
 
\bibitem[Hatchell, Fuller, \& Ladd 1999]{hat99} Hatchell, J., Fuller,
G.A., Ladd, E.F., 1999, A\&A, 346, 278

\bibitem[Hogerheijde et al. 1998]{hog98} Hogerheijde, M.R., van Dishoeck, 
E.F., Blake, G.A., van Langevelde, H.J., 1998, ApJ, 502, 315

\bibitem[Hogerheijde et al. 1999]{hog99} Hogerheijde, M.R., van Dishoeck, 
E.F., Salverda, J.M., Blake, G.A., 1999, ApJ, 513, 350

\bibitem[Irvine, Goldsmith, \& Hjalmarson 1987]{irv87} Irvine, W.M., 
Goldsmith, P.F., Hjalmarson, \AA, 1987, IAU Symp. No. 120, Interstellar
Processes, eds. D.J. Hollenbach and H.A., Thronson (Dordrecht: Reidel), 
p. 561

\bibitem[Kitamura et al. 1997]{kit97} Kitamura, Y., Saito, M., Kawabe, R.,
Sunada, K., 1997, IAU Symp. No. 182, Herbig-Haro Flows and the Birth of 
Low Mass Stars, eds. B. Reipurth and C. Bertout, (Kluwer: Netherland), 
p. 381

\bibitem[Ladd et al. 1991]{lad91} Ladd, E.F., Adams, F.C., Casey, S.,
Davidson, J.A., Fuller, G.A., Harper, D.A., Myers, P.C., Padman, R.,
1991, ApJ, 366, 203

\bibitem[Langer, Velusamy, \& Xie 1996]{lan96} Langer, W.D., Velusamy, T.,
Xie, T., 1996, ApJ, 468, L41


\bibitem[Leung 1978]{leu78} Leung, C.M., 1978, ApJ, 427, 441

\bibitem[Leung \& Brown 1977]{leu77} Leung, C.M., Brown, R.L., 1977, ApJL, 
214, L73

\bibitem[Li \& Shu 1996]{li96} Li, Z.-Y., Shu, F.H., 1996, ApJ, 468, 261

\bibitem[Madden et al. 1989]{mad98} Madden, S.C., Irvine, W.M., 
Matthews, H.E., Friberg, P., Swade, D.A., 1989, AJ, 97, 1403

\bibitem[Mardones 1998]{mar98} Mardones, D., 1998, Ph.D. thesis, Harvard 
University
 
\bibitem[Mardones et al. 1997]{mar97} Mardones, D., Myers, P.C., 
Tafalla, M., Wilner, D.J., Bachiller, R., Garay, G., 1997, ApJ, 489, 719

\bibitem[Masson \& Chernin 1993]{mas93} Masson, C.R., Chernin, L.M.,
1993, ApJ, 414, 230


\bibitem[Momose et al. 1998]{mom98} Momose, M., Ohashi, N., Kawabe, R.,
Nakano, T., Hayashi, M., 1998, ApJ, 504, 314

\bibitem[Motte, Andr\'e, \& Neri 1998]{mot98} Motte, F., Andr\'e, P., \&
Neri, R., 1998, A\&A, 336, 150
 
\bibitem[Myers et al. 1995]{mye95} Myers, P.C., Bachiller, R.,
Caselli, P., Fuller, G.A., Mardones, D., Tafalla, M., Wilner, D.J.,
1995, ApJ, 449, L65

\bibitem[Ohashi et al. 1997]{oha97} Ohashi, N., Hayashi, M., Ho, P.T.P.,
Momose, M., 1997, ApJ, 475, 211

\bibitem[Park, Kim, \& Minh 1999]{par99} Park, Y.-S., Kim, J., Minh, Y.C., 
1999, ApJ, 520, 223
 
\bibitem[Parker, Padman, \& Scott 1991]{par91} Parker, N.D., Padman, R.,
Scott, P.F., 1991, MNRAS, 252, 442

\bibitem[Rawlings 1996]{raw96} Rawlings, J.M., 1996, ApSS, 237, 299

\bibitem[Sandell et al. 1994]{san94} Sandell, G., Knee, L.B.G., Aspin,
C., Robson, I.E., Russell, A.P.G., 1994, ApJL, 285, 1
 
\bibitem[Tafalla \& Myers 1997]{taf97} Tafalla, M., Myers, P.C., 1997, 
ApJ, 491, 653
 
\bibitem[van Dishoeck \& Blake 1998]{van98} van Dishoeck, F.F., Blake, 
G.A., 1998, ARAA, 36, 317

\bibitem[Yang et al. 1997]{yan97} Yang, J., Ohashi, N., Yan, J., Liu, C.,
Kaifu, N., Kimura, H., 1997, ApJ, 475, 683


\bibitem[Zhou et al. 1993]{zho93} Zhou, S., Evans, N.J.II., K\"ompe, C., 
Walmsley, C.M., 1993, ApJ, 404, 232


\end{thebibliography}
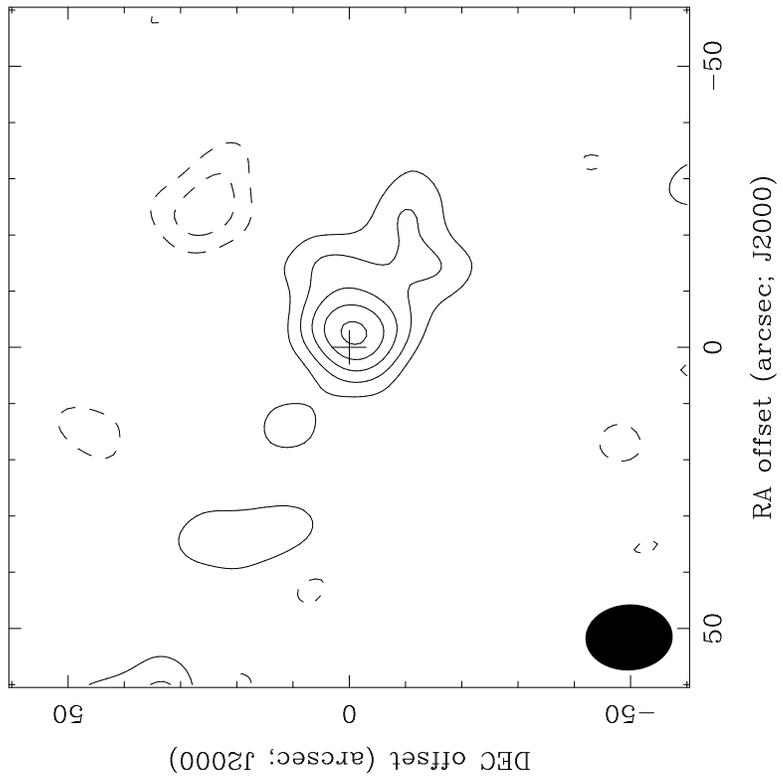
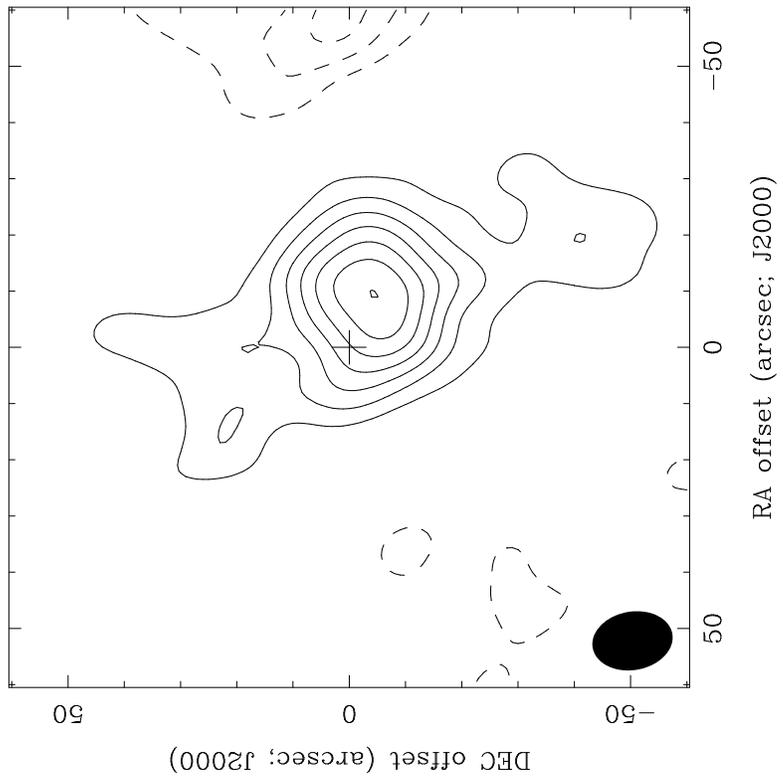
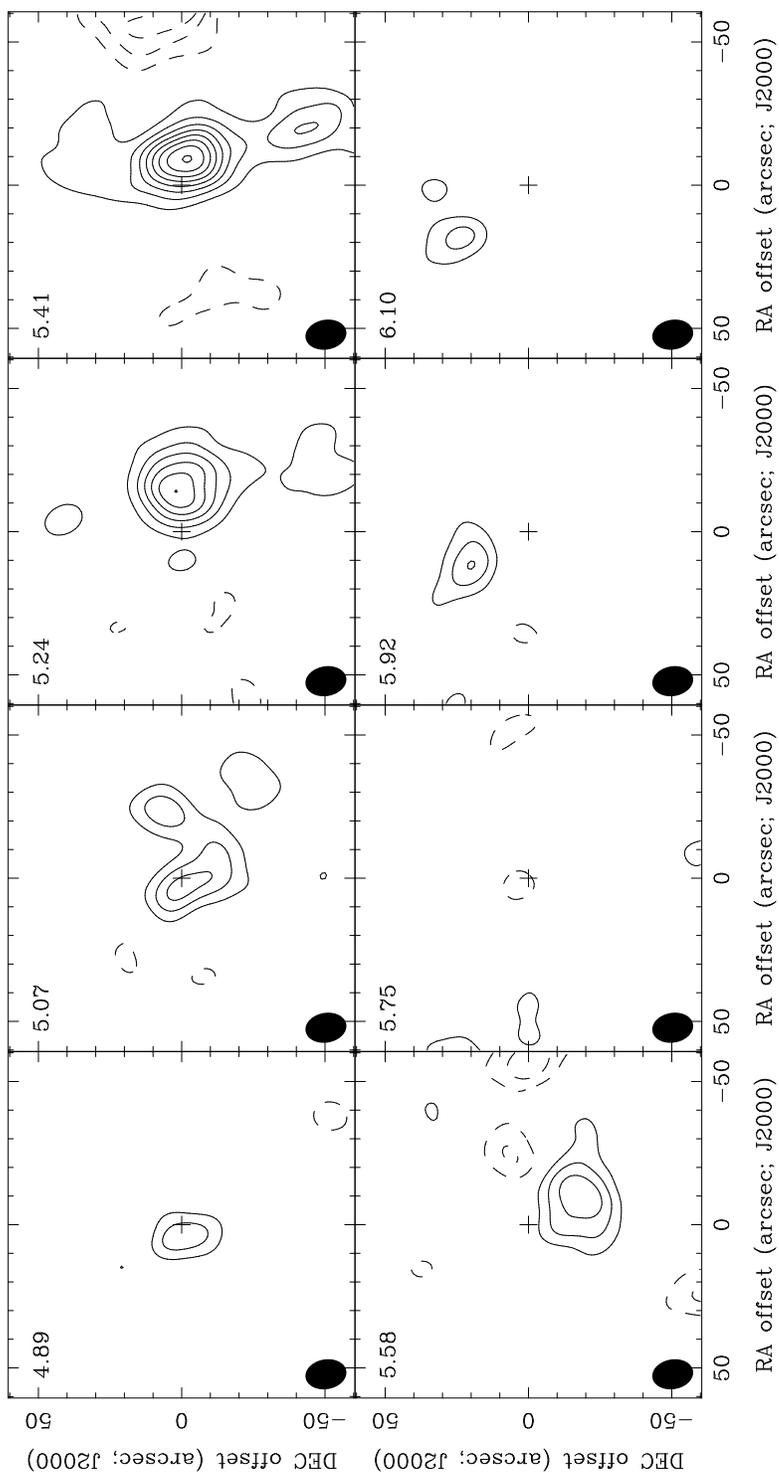
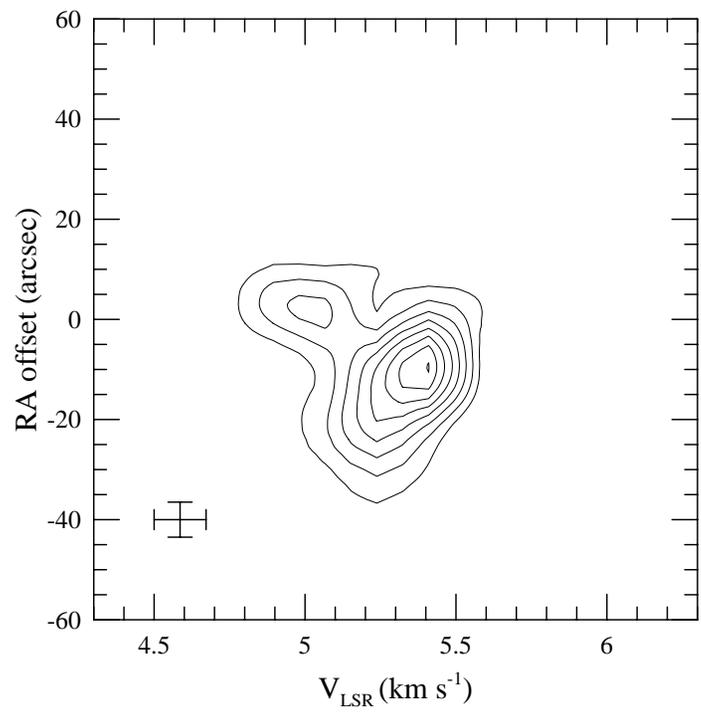
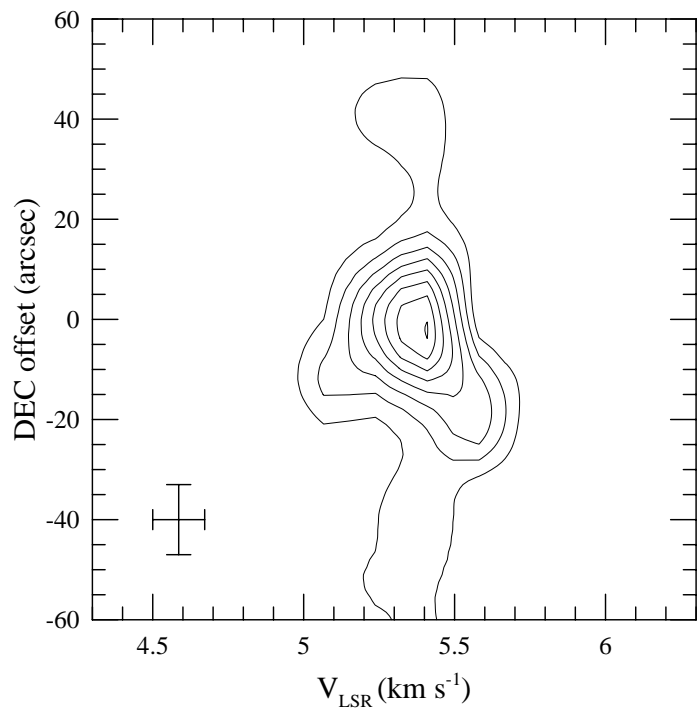
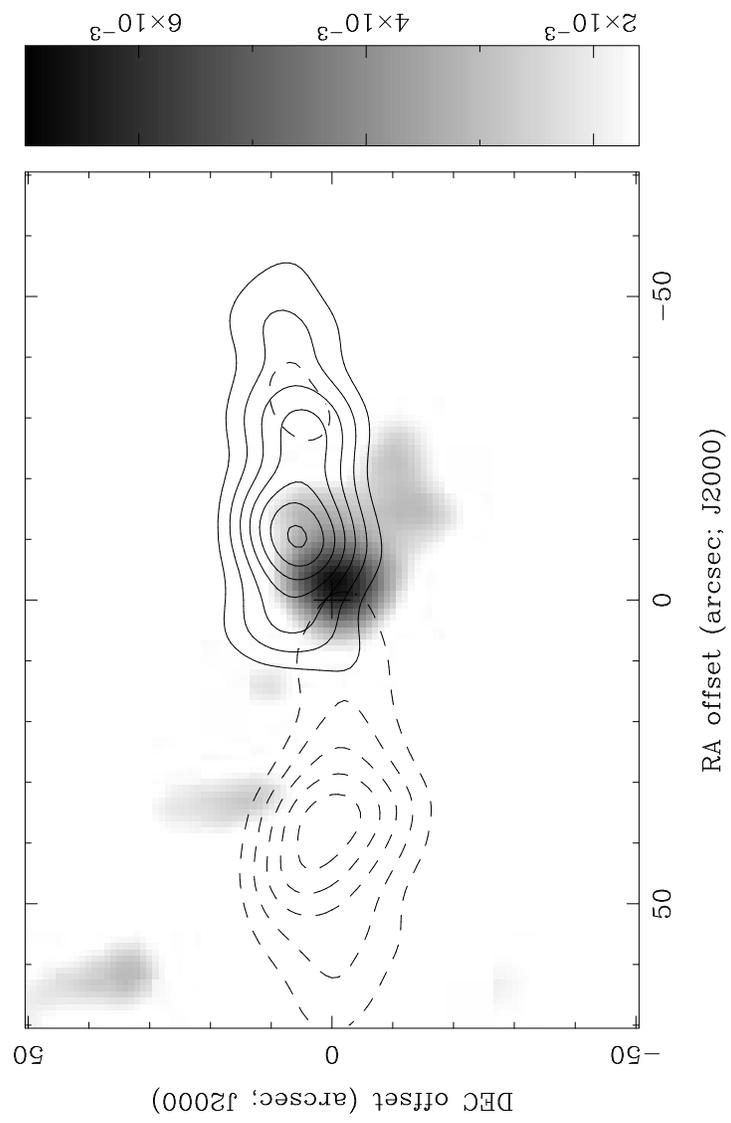
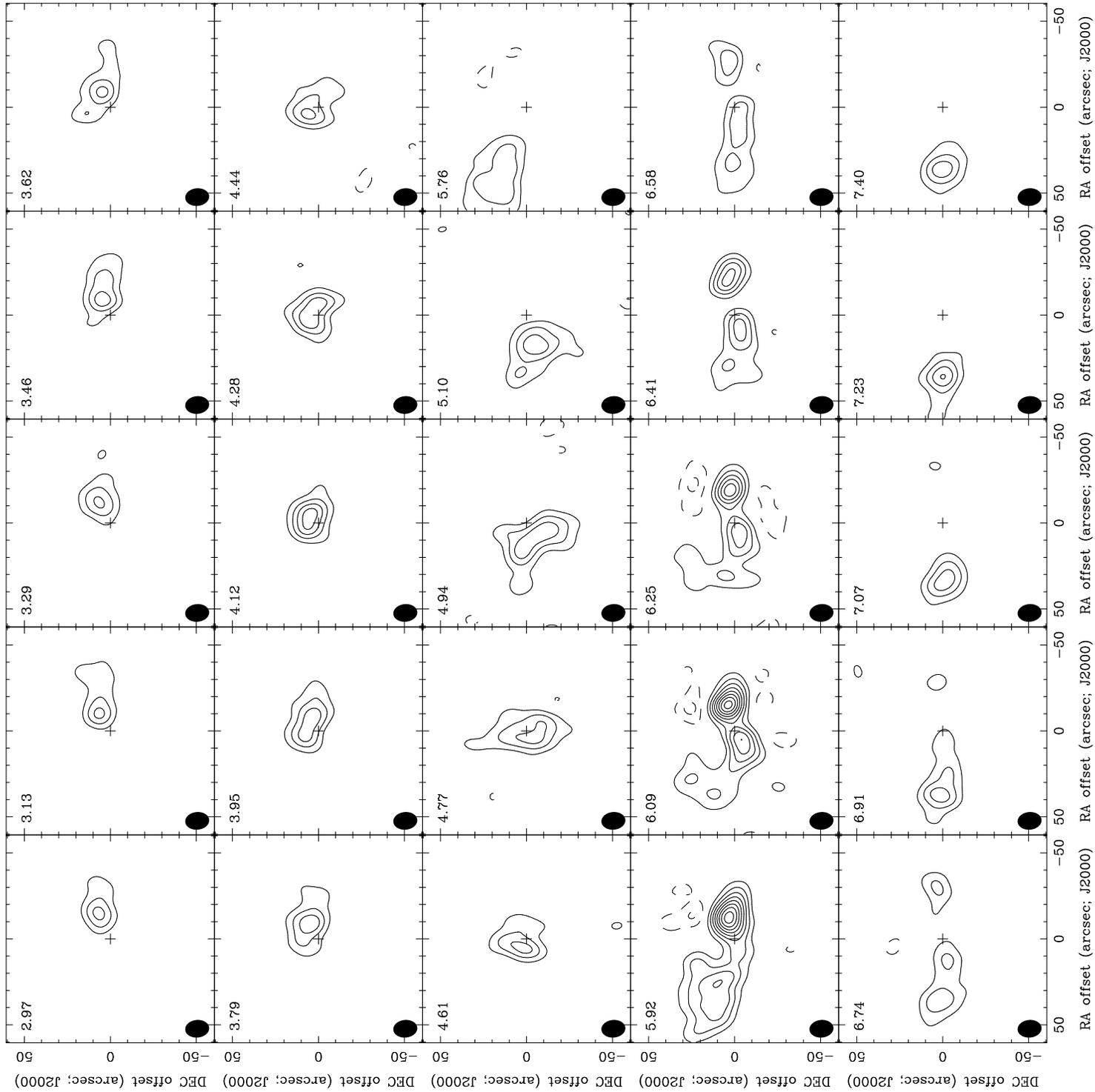
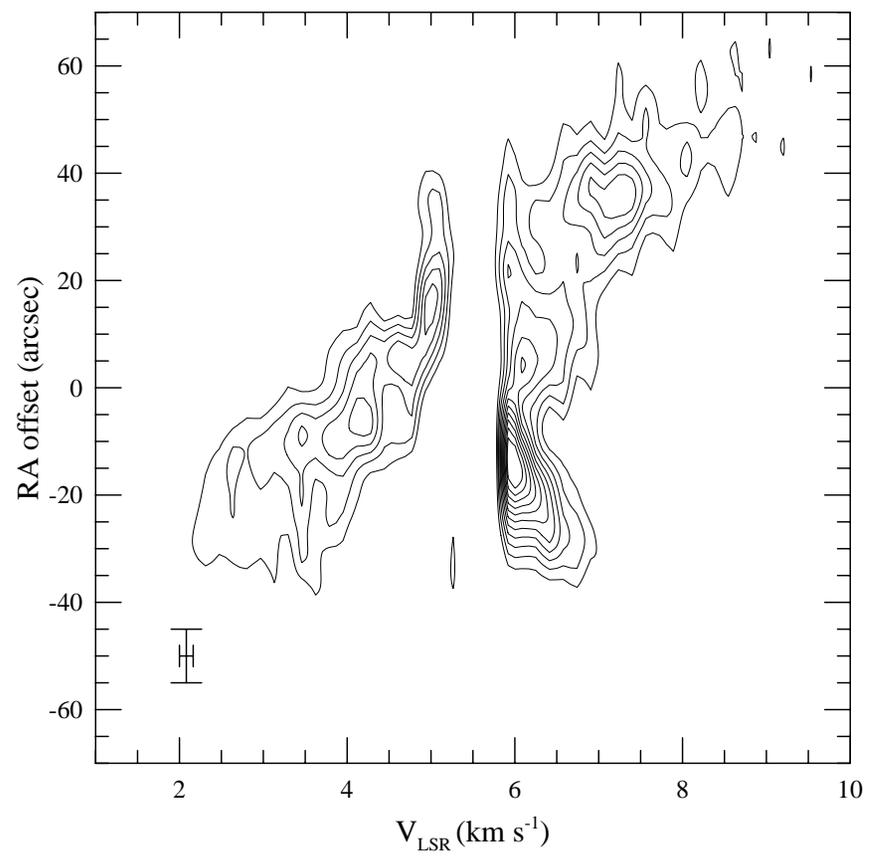
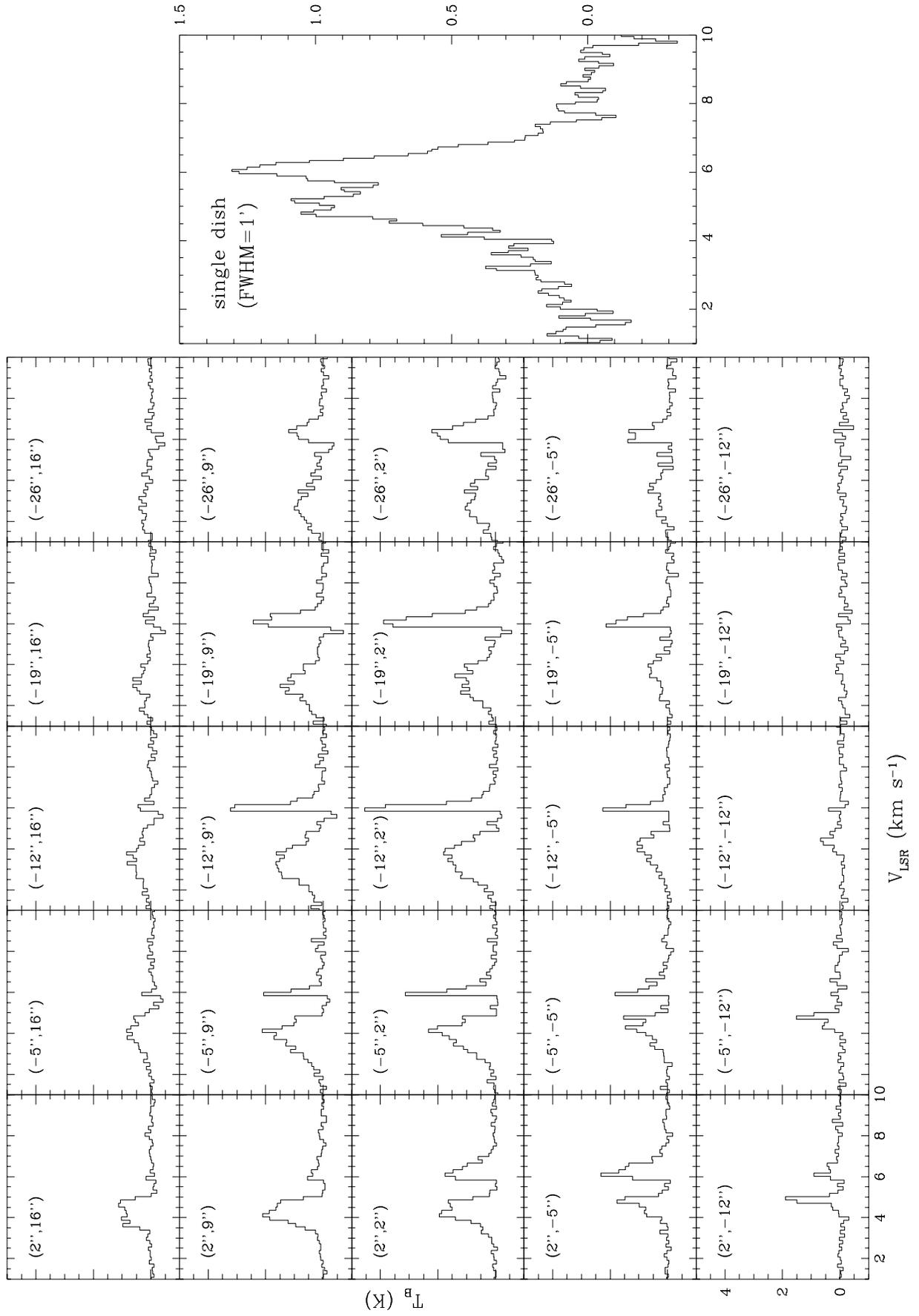
\end{document}